\documentclass[pra,onecolumn,a4paper,superscriptaddress,showpacs,showkeys,preprintnumbers,12pt]{revtex4}
\usepackage{graphicx}
\usepackage{amsmath}
\usepackage{amsfonts}
\usepackage{amssymb}
\usepackage{bbm}
\usepackage{pst-plot}
\usepackage{pst-node}
\usepackage{pst-coil}
\newcommand{\tr}{{\rm{Tr}}}

\begin{document}

\preprint{\begin{tabular}{r}
UWThPh-2005-16\\
December 2005\\
\end{tabular}}

\title{Decoherence modes of entangled qubits within neutron interferometry}

\author{Reinhold A. Bertlmann}
\author{Katharina Durstberger}
\email{katharina.durstberger@univie.ac.at}
\affiliation{Institute for Theoretical Physics, University of Vienna\\
Boltzmanngasse 5, 1090 Vienna, Austria}
\author{Yuji Hasegawa}
\affiliation{Atominstitut der \"Osterreichischen Universit\"aten\\
Stadionallee 2, 1020 Vienna, Austria} \affiliation{PRESTO, Japan Science and Technology Agency\\
4-1-8 Honcho Kawaguchi, Saitama, Japan\\\vspace{1cm}}


\begin{abstract}
We study two different decoherence modes for entangled qubits by considering a Liouville -- von
Neumann master equation. Mode A is determined by projection operators onto the eigenstates of the
Hamiltonian and mode B by projectors onto rotated states. We present solutions for general and for
Bell diagonal states and calculate for the later the mixedness and the amount of entanglement
given by the concurrence.

We propose a realization of the decoherence modes within neutron interferometry by applying
fluctuating magnetic fields. An experimental test of the Kraus operator decomposition describing
the evolution of the system for each mode is presented.
\end{abstract}

\pacs{03.65.Yz, 03.75.Dg, 42.50.-p}

\keywords{Entanglement, decoherence, master equation, Kraus operators, Bell diagonal states,
neutron interferometry}

\maketitle

\section{Introduction}

Closed quantum systems are idealizations which do not exist in a real physical world. Actually,
one always has to deal with open quantum systems which arise due to an interaction of the system
under consideration with an external environment (e.g. reservoir, heat bath)
\cite{BreuerPetruccione,NielsenChuang,GiuliniJoosKieferKupschStamatescuZeh}. The system --
environment interaction causes a phenomenon known as decoherence: quantum correlations and
interferences are destroyed in course of time; the system shows more and more classical behavior.
The theory of decoherence is one candidate to solve the question why our world looks so classical
\cite{GiuliniJoosKieferKupschStamatescuZeh,Zurek1991}.

The total Hamiltonian of system and environment generates a unitary time evolution $U(t)$ and is
of the form $H_{SE}=H\otimes\mathbbm{1}_E+\mathbbm{1}\otimes H_E+H_I$, where $H$, $H_E$ and $H_I$
are, respectively, the system, environment and interaction Hamiltonians. The evolution of the
system, represented by the density matrix $\rho(t)$, or the reduced dynamics is obtained by
tracing over the environmental degrees of freedom $\rho(t)=\tr_E \,\rho_{S+E}(t)=\tr_E (U(t)
\rho_{S+E}(0)U^{\dagger}(t))$ and thus inheriting a nonunitary evolution for the system in contrast
to closed systems.

In most of the cases we do not have access or information about the dynamics of the environment.
Therefore we have to describe the evolution of the system by an effective dynamics: the Liouville
-- von Neumann master equation. Thereby it is not so important to know the exact Hamiltonian and
the nature of the environment but only its effects on the system. Our strategy in this paper is to
propose several effective models which do not care about the exact nature of decoherence but
provide scenarios how decoherence can affect a system.

Under several assumptions \cite{BreuerPetruccione}, such as Markovian semigroup approach, complete
positivity, initial decoupling of system and environment, and weak coupling, the dynamics of the
system can be described by a Liouville -- von Neumann master equation
\begin{equation}\label{time-evolution}
    \frac{\partial}{\partial t}\rho(t)=-i[H(t),\rho(t)]-D[\rho(t)]\;.
\end{equation}
Lindblad and Gorini--Kossakowski--Sudarshan \cite{Lindblad, GoriniKossakowskiSudarshan} derived
the most general structure of the dissipator
\begin{equation}\label{dissipator-allg}
    D[\rho(t)]=\frac{1}{2}\sum_{k} \bigl(A_k^{\dag}A_k \rho(t) +
    \rho(t)A_k^{\dag}A_k-2A_k\rho(t)A_k^{\dag}\bigr)\,,
\end{equation}
where $A_k$ represents a so-called Lindblad generator. The sum is taken over an arbitrary number
of components but maximally up to $n^2-1$, where $n$ denotes the dimension of the system. For
simplicity we choose the generators to be projectors such that $A_k=\sqrt{\lambda_k}P_k$ with
$P_k^2=P_k$ (see Ref.\cite{BertlmannGrimus2002}) which gives for the dissipator
\begin{equation}\label{dissipator}
    D[\rho]=\frac{1}{2}\sum_k \lambda_k\bigl(P_k \rho +
    \rho P_k-2P_k\rho P_k\bigr)\;.
\end{equation}

The paper is organized as follows. In the next section we introduce and discuss two possible
decoherence scenarios for a two qubit system by choosing different projection operators $P_k$. In
Sect.\ref{sect.bell-diag} the two decoherence modes are discussed for the special case of Bell
diagonal states. In Sect.\ref{sect.expI} we propose a realization of the decoherence modes within
neutron interferometry via random magnetic fields which represent the environment. In
Sect.\ref{sect.expII} we present the Kraus operator decomposition. The action of this
decomposition is mathematically equivalent to the Lindblad form of the Liouville -- von Neumann
equation. We can test this equivalence by a simple experiment with single neutrons.

\section{Decoherence modes in a two qubit system}\label{sect.deco-modes}

Let us consider a two qubit system with Hilbertspace
$\mathcal{H}=\mathcal{H}^{(1)}\otimes\mathcal{H}^{(2)}=\mathbb{C}^2\otimes\mathbb{C}^2$
where $\{\lvert e_k\rangle\}_{k=1,\ldots, 4}$ denotes an eigenbasis defined by $H\lvert
e_k\rangle=E_k\lvert e_k\rangle$, with $H=H^{(1)}\otimes\mathbbm{1}+\mathbbm{1}\otimes
H^{(2)}$ the Hamiltonian of the undisturbed system. A general state $\rho$ of the system
can be expressed in the eigenbasis $\rho=\sum_{k,j} \rho_{kj}\lvert e_k\rangle\langle
e_j\rvert$, where $(\rho_{kj})$ denotes the $4\times4$ coefficient matrix.

We consider Lindblad generators $P_k$ that project onto one-dimensional subspaces and fulfil
$\sum_k P_k=\mathbbm{1}$, furthermore we assume that only one dissipation parameter $\lambda$
parameterizes the strength of the interaction and therefore of the decoherence. Then the
dissipator, Eq.\eqref{dissipator}, can be written as
\begin{equation}\label{4dim.diss.ansatz1}
    D[\rho]=\lambda\bigl(\rho- \sum_{k=1}^{4} P_k\rho P_k\bigr)\;.
\end{equation}

In the following sections we solve the Liouville -- von Neumann equation \eqref{time-evolution}
with the dissipator \eqref{4dim.diss.ansatz1} by assuming different projection operators $P_k$,
what we call decoherence modes.

\subsection{Mode A}

The first mode describes the simplest possible case. The Lindblad generators are chosen to be
projectors $P_k=\lvert e_k\rangle\langle e_k\lvert$ onto the eigenbasis of the Hamiltonian (see
Refs.\cite{BertlmannDurstbergerHiesmayr2002,BertlmannGrimus2001}). In this mode of decoherence the
time evolution \eqref{time-evolution} for the coefficient matrix is given by
\begin{equation}
\begin{split}
    \dot{\rho_{kj}}&=\bigl(-i(E_k-E_j)-\lambda_A\bigr)\rho_{kj}\qquad\mbox{for}\; k\neq j\\
    \dot{\rho_{kk}}&=0\;,
\end{split}
\end{equation}
which can be easily solved
\begin{equation}\label{timeevol.4dim1}
\begin{split}
    \rho_{kj}(t)&=e^{-i(E_k-E_j)t}e^{-\lambda_A t}\rho_{kj}(0)\qquad\mbox{for}\; k\neq j\\
    \rho_{kk}(t)&=\rho_{kk}(0)\;.
\end{split}
\end{equation}
The decoherence affects only the off-diagonal elements and leaves the diagonal elements
untouched.

\subsection{Mode B}

For the second mode the Lindblad generators are chosen to be projectors
$\widetilde{P_k}=\lvert\widetilde{e_k}\rangle\langle\widetilde{e_k}\lvert$ onto the
following states
\begin{equation}\label{proj.states}
    \lvert\widetilde{e_{1,3}}\rangle
    =\frac{1}{\sqrt{2}}(\lvert e_1\rangle \pm\lvert e_3\rangle)\;,\hspace{2cm}
    \lvert\widetilde{e_{2,4}}\rangle
    =\frac{1}{\sqrt{2}}(\lvert e_2\rangle \pm\lvert e_4\rangle)\;,
\end{equation}
where the upper (lower) sign corresponds to the first (second) index.

The time evolution \eqref{time-evolution} of the coefficient matrix can be separated into 3 types
of differential equations. Type I is valid for the components
$\rho_{12},\rho_{14},\rho_{23},\rho_{34}$ and has the structure
\begin{equation}\label{block1}
    \dot{\rho_{12}}=\bigl(-i(E_1-E_2)-\lambda_B\bigr)\rho_{12}\;,
\end{equation}
in analogy to mode A. Type II holds for the diagonal components and reveals pairwise coupled
differential equations for $\rho_{11}$ -- $\rho_{33}$ and $\rho_{22}$ -- $\rho_{44}$ of the form
\begin{equation}\label{block2}
    \dot{\rho_{11}}=-\frac{\lambda_B}{2}\rho_{11}+\frac{\lambda_B}{2}\rho_{33}\;,\hspace{2.3cm}
    \dot{\rho_{33}}=\frac{\lambda_B}{2}\rho_{11}-\frac{\lambda_B}{2}\rho_{33}\;.
\end{equation}
Type III also gives pairwise coupled differential equations
\begin{equation}\label{block3}
    \dot{\rho_{13}}=\bigl(-i(E_1-E_3)-\frac{\lambda_B}{2}\bigr)\rho_{13}+\frac{\lambda_B}{2}\rho_{31}\;,\hspace{1cm}
    \dot{\rho_{31}}=\frac{\lambda_B}{2}\rho_{13}+\bigl(i(E_1-E_3)-\frac{\lambda_B}{2}\bigr)\rho_{31}\;,
\end{equation}
valid for the components $\rho_{13}$ -- $\rho_{31}$ and $\rho_{24}$ -- $\rho_{42}$. The solutions
for the several
types of differential equations are:\\
for type I, Eq.\eqref{block1},
\begin{equation}\label{timeevol.4dim2a}
    \rho_{12}(t)=e^{-i(E_1-E_2)t}e^{-\lambda_B t}\rho_{12}(0)\;,
\end{equation}
for type II, Eq.\eqref{block2},
\begin{equation}\label{timeevol.4dim2b}
\begin{split}
    \rho_{11}(t)&=\frac{1}{2}(1+e^{-\lambda_B t})\rho_{11}(0)+\frac{1}{2}(1-e^{-\lambda_B t})\rho_{33}(0)\;,\\
    \rho_{33}(t)&=\frac{1}{2}(1-e^{-\lambda_B t})\rho_{11}(0)+\frac{1}{2}(1+e^{-\lambda_B
    t})\rho_{33}(0)\;,
\end{split}
\end{equation}
for type III, Eq.\eqref{block3},
\begin{equation}\label{timeevol.4dim2c}
\begin{split}
    \rho_{13}(t)&=e^{-\frac{\lambda_B t}{2}}\biggl(
    \Bigl(\cosh\frac{\mu t}{2}-\frac{2i(E_1-E_3)}{\mu}\sinh\frac{\mu t}{2}\Bigr)\rho_{13}(0)
    +\frac{\lambda_B}{\mu}\sinh\frac{\mu t}{2}\rho_{31}(0)\biggr)\;,\\
    \rho_{31}(t)&=e^{-\frac{\lambda_B t}{2}}\biggl(
    \Bigl(\cosh\frac{\mu t}{2}+\frac{2i(E_1-E_3)}{\mu}\sinh\frac{\mu t}{2}\Bigr)\rho_{31}(0)
    +\frac{\lambda_B}{\mu}\sinh\frac{\mu t}{2}\rho_{13}(0)\biggr)\;,
\end{split}
\end{equation}
where $\mu=\sqrt{\lambda_B^2-4(E_1-E_3)^2}$.\\

Decoherence mode B affects not only the off-diagonal elements of the density matrix but also the
diagonal ones.

In the case we consider in this paper -- decoherence modes in neutron interferometry,
Sect.\ref{sect.expI} -- the system is given by the free neutron passing through the
interferometer, whereas the magnetic fields placed in represent the external environment. Thus
there is no splitting in the energies, e.g., $E_1=E_3$, $ E_2=E_4$ and $\mu=\lambda_B$, so that
type III is equal to type II.\\

\emph{Remark.} It is worth noting here that the choice of projection states,
Eq.\eqref{proj.states}, corresponds to a rotation of states in one subspace. Suppose we split the
eigenstates of the undisturbed Hamiltonian in eigenstates of the subspace Hamiltonians $\{\lvert
a_1\rangle, \lvert a_2\rangle\}$ and $\{\lvert b_1\rangle, \lvert b_2\rangle\}$ in the following
way
\begin{equation}
    \lvert e_{1,3}\rangle=\lvert a_{1,2}\rangle\lvert b_1\rangle\;, \hspace{2cm}
    \lvert e_{2,4}\rangle=\lvert a_{1,2}\rangle\lvert b_2\rangle\;.
\end{equation}
Now consider a rotation of the first subbasis, $\{\lvert
+\rangle=\frac{1}{\sqrt{2}}(\lvert a_1\rangle +\lvert a_2\rangle), \lvert
-\rangle=\frac{1}{\sqrt{2}}(\lvert a_1\rangle -\lvert a_2\rangle)\}$, the second subbasis
is left untouched. The basis of the total Hilbertspace changes
\begin{equation}
\begin{split}
    &\lvert\widetilde{e_1}\rangle=\lvert +\rangle\lvert b_1\rangle
    =\frac{1}{\sqrt{2}}(\lvert e_1\rangle +\lvert e_3\rangle)\;,\hspace{2cm}
    \lvert\widetilde{e_2}\rangle=\lvert +\rangle\lvert b_2\rangle
    =\frac{1}{\sqrt{2}}(\lvert e_2\rangle +\lvert e_4\rangle)\;,\\
    &\lvert\widetilde{e_3}\rangle=\lvert -\rangle\lvert b_1\rangle
    =\frac{1}{\sqrt{2}}(\lvert e_1\rangle -\lvert e_3\rangle)\;,\hspace{2cm}
    \lvert\widetilde{e_4}\rangle=\lvert -\rangle\lvert b_2\rangle
    =\frac{1}{\sqrt{2}}(\lvert e_2\rangle -\lvert e_4\rangle)\;,
\end{split}
\end{equation}
which corresponds exactly to the states used in Eq.\eqref{proj.states}. Therefore decoherence mode
B can be denoted ``$R\otimes E$\,'' to indicate the rotation of the first subspace and the
untouched eigenbasis in the second subspace whereas mode A can be labelled by
``$E\otimes E$\,''.\\

In the case of photons (see, e.g., \cite{BouwmeesterEkertZeilinger}) the eigenbasis $E$
corresponds to horizontal $\lvert H\rangle$ and vertical $\lvert V\rangle$ polarization whereas
the rotated basis $R$ represents polarization states $\lvert+45^\circ\rangle$ and
$\lvert-45^\circ\rangle$. In the case of neutral kaons (for an overview see, e.g.,
\cite{HiesmayrDiss,BertlmannSchladming}) we can identify the eigenbasis $E$ with the short- and
long-lived states $\lvert K_S\rangle$ and $\lvert K_L\rangle$ and the rotated basis $R$ with
$\lvert K^0\rangle$ and $\lvert \bar K^0\rangle$.

\section{Initial conditions -- Bell diagonal states}\label{sect.bell-diag}

We want to illustrate the above discussed decoherence modes by choosing a certain class of states
as initial conditions -- the so-called Bell diagonal states $\rho=\sum_i \nu_i
\lvert\Psi_i\rangle\langle\Psi_i\rvert$ with $\sum_i \nu_i=1$, which are diagonal in the Bell
basis
\begin{equation}
    \lvert \Psi_{1,2}\rangle=\frac{1}{\sqrt{2}}(\lvert e_1\rangle\pm\lvert
    e_4\rangle)\;,\hspace{2cm}
    \lvert \Psi_{3,4}\rangle=\frac{1}{\sqrt{2}}(\lvert e_2\rangle\pm\lvert
    e_3\rangle)\;.
\end{equation}
In the standard basis they are expressed by
\begin{equation}\label{Bell-diag.initial}
    \rho=\frac{1}{2}\begin{pmatrix}
      \nu_1+\nu_2 & 0 & 0 & \nu_1-\nu_2 \\
      0 & \nu_3+\nu_4 & \nu_3-\nu_4 & 0 \\
      0 & \nu_3-\nu_4 & \nu_3+\nu_4 & 0 \\
      \nu_1-\nu_2 & 0 & 0 & \nu_1+\nu_2 \\
    \end{pmatrix}=\frac{1}{2}\begin{pmatrix}
      \Sigma_1 & 0 & 0 & \Delta_1 \\
      0 & \Sigma_2 & \Delta_2 & 0 \\
      0 & \Delta_2 & \Sigma_2 & 0 \\
      \Delta_1 & 0 & 0 & \Sigma_1 \\
    \end{pmatrix}\;,
\end{equation}
using the notation $\Sigma_1=\nu_1+\nu_2$, $\Sigma_2=\nu_3+\nu_4$, $\Delta_1=\nu_1-\nu_2$,
$\Delta_2=\nu_3-\nu_4$.

States are characterized by two quantities: mixing and entanglement. The mixedness, defined as
$\delta=\tr \rho^2$, ranges between $1$ (pure states) and $\frac{1}{4}$ (maximally mixed states)
and is given by $\delta=\nu_1^2+\nu_2^2+\nu_3^2+\nu_4^2$ for Bell diagonal states. The concurrence
$C$ \cite{HillWootters,Wootters1997,Wootters2001} is a suitable quantity that measures the
entanglement contained in a state $\rho$. It is defined by
$C(\rho)=\max\{0,\mu_1-\mu_2-\mu_3-\mu_4\}$, where $\mu_i$ are the square roots of the eigenvalues
in decreasing order of the matrix
$R=\rho\,(\sigma_y\otimes\sigma_y)\,\rho^{\ast}\,(\sigma_y\otimes\sigma_y)$ and $\rho^{\ast}$
denotes complex conjugation in the standard basis. The concurrence varies between $1$ (maximally
entangled states) and $0$ (separable states) and is given by
$C=\max\bigl\{0,2\max\{\nu_i\}-1\bigr\}$ for Bell diagonal states, depending on which weight
$\nu_i$ is the largest. A Bell diagonal state can only be entangled ($C>0$) if the largest
eigenvalue fulfills $\nu_i>\frac{1}{2}$.

The special case of a pure and maximally entangled Bell state, e.g., the Bell singlet state
$\lvert\Psi_4\rangle$, where $\nu_4=1$ and $\nu_1=\nu_2=\nu_3=0$ or $\Sigma_1=\Delta_1=0$ and
$\Sigma_2=-\Delta_2=1$, yields $\delta=1$ and $C=1$.

\subsection{Mode A}

The initial Bell diagonal state, Eq.\eqref{Bell-diag.initial}, evolves in time according to mode
A, Eq.\eqref{timeevol.4dim1}, into the state
\begin{equation}
    \rho(t)=\frac{1}{2}\begin{pmatrix}
      \Sigma_1 & 0 & 0 & e^{-\lambda_A t}\Delta_1 \\
      0 & \Sigma_2 & e^{-\lambda_A t}\Delta_2 & 0 \\
      0 & e^{-\lambda_A t}\Delta_2 & \Sigma_2 & 0 \\
      e^{-\lambda_A t}\Delta_1 & 0 & 0 & \Sigma_1 \\
    \end{pmatrix}\;.
\end{equation}
For the mixedness of the state we get $\delta =\frac{1}{2}\bigl(\Sigma_1^2+\Sigma_2^2+
e^{-2\lambda_A t}(\Delta_1^2+\Delta_2^2))$ and for the concurrence we find
$C(\rho)=\max\bigl\{0,2\max\{\mu_i\}-1\bigr\}$, where $\mu_{1,2}=\frac{1}{2}(\Sigma_1\pm
e^{-\lambda_A t}\Delta_1)$ and $\mu_{3,4}=\frac{1}{2}(\Sigma_2\pm e^{-\lambda_A t}\Delta_2)$.

Choosing the Bell singlet state $\lvert\Psi_4\rangle$ the mixedness $\delta
=\frac{1}{2}(1+e^{-2\lambda_A t})$ ranges from a pure state ($\delta=1$) to a mixed but not
maximally mixed state ($\delta\xrightarrow[]{t\rightarrow\infty}\frac{1}{2}$). The concurrence
$C(\rho)=e^{-\lambda_A t}$ decreases exponentially from a maximally entangled state ($C=1$) to an
asymptotically separable state ($C\xrightarrow[]{t\rightarrow\infty} 0$). The behavior of $\delta$
and $C$ is plotted in Fig.\ref{fig.comparison}.

\subsection{Mode B}

The second mode, Eqs.\eqref{timeevol.4dim2a}-\eqref{timeevol.4dim2c}, generates the density matrix
\begin{equation}
    \rho(t)=\frac{1}{4}
    \begin{pmatrix}
      1-e^{-\lambda_B t}\,\Delta & 0 & 0 & 2e^{-\lambda_B t}\Delta_1 \\
      0 & 1+e^{-\lambda_B t}\,\Delta & 2e^{-\lambda_B t}\Delta_2 & 0 \\
      0 & 2e^{-\lambda_B t}\Delta_2 & 1+e^{-\lambda_B t}\,\Delta & 0 \\
      2e^{-\lambda_B t}\Delta_1 & 0 & 0 & 1-e^{-\lambda_B t}\,\Delta \\
    \end{pmatrix}\;,
\end{equation}
with the notation $\Delta=\Sigma_1-\Sigma_2$.

We obtain for the mixedness $\delta=\frac{1}{4}\Bigl(1+e^{-2\lambda_B
t}\bigl(2\Delta_1^2+2\Delta_2^2+(\Sigma_1+\Sigma_2)^2\bigr)\Bigr)$ and for the entanglement
$C(\rho)=\max\bigl\{0,2\max\{\mu_i\}-1\bigr\}$, where $\mu_{1,2}=\frac{1}{4}(1+e^{-\lambda_B
t}(\Delta\pm2\Delta_2))$ and $\mu_{3,4}=\frac{1}{4}(1-e^{-\lambda_B t}(\Delta\mp2\Delta_1))$.

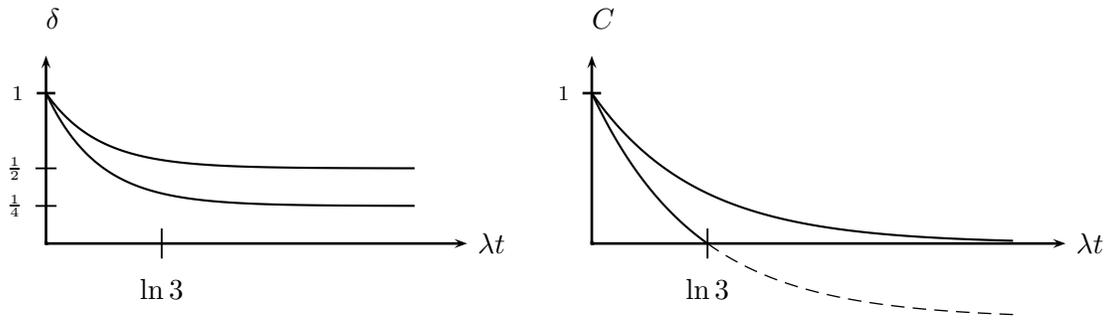
\begin{figure}[htbp]
\centering
\begin{pspicture}(-0.5,-0.8)(5,3.3)
\psset{xunit=1.4cm,yunit=2cm}
{\scriptsize\psaxes[linewidth=1pt,ticks=y,labels=y,showorigin=false]{->}(0,0)(0,0)(4,1.25)}
\psplot[plotstyle=curve,plotpoints=200]{0}{3.5}{0.5 2.71828 2 neg x mul exp 2 div add}
\psplot[plotstyle=curve]{0}{3.5}{1 4 div 2.71828 2 neg x mul exp 3 mul 4 div add}
\rput[l](0,1.5){$\delta$}%
\rput[l](4.1,0){$\lambda t$}%
\psline[linewidth=0.7pt](1.0986,-0.1)(1.0986,0.1)
\rput(1.0986,-0.3){$\ln3$}
\psline[linewidth=0.7pt](-0.1,0.5)(0.1,0.5)
\rput(-0.3,0.5){\tiny $\frac{1}{2}$}
\psline[linewidth=0.7pt](-0.1,0.25)(0.1,0.25)
\rput(-0.3,0.25){\tiny $\frac{1}{4}$}
\end{pspicture}
\hspace{1.5cm}
\begin{pspicture}(-0.5,-0.8)(6,3.3)
\psset{xunit=1.4cm,yunit=2cm}
{\scriptsize\psaxes[linewidth=1pt,ticks=y,labels=y,showorigin=false]{->}(0,0)(0,0)(4.5,1.25)}
\psplot[plotstyle=curve]{0}{4}{2.71828 x neg exp}
\psplot[plotstyle=curve]{0}{1.0986}{1 neg 2 div 2.71828 x neg exp 3 mul 2 div add}
\psplot[plotstyle=curve,linestyle=dashed,linewidth=0.5pt]{1.0986}{4}{1 neg 2 div 2.71828 x neg exp 3 mul 2 div add}%
\rput[l](0,1.5){$C$}%
\rput[l](4.6,0){$\lambda t$}%
\psline[linewidth=0.7pt](1.0986,-0.1)(1.0986,0.1)
\rput(1.0986,-0.3){$\ln3$}
\end{pspicture}
  \caption{Graphical comparison of the mixedness $\delta$ and the concurrence $C$ in dependence of
  $\lambda t$ for mode A and B for the Bell singlet state $\lvert\Psi_4\rangle$.
  The upper curves correspond to mode A and the lower ones to mode B.}\label{fig.comparison}
\end{figure}

The special case of $\lvert\Psi_4\rangle$ yields the following results. The mixedness
$\delta=\frac{1}{4}(1+3e^{-2\lambda_B t})$ varies from a pure state ($\delta=1$) to a maximally
mixed state ($\delta\xrightarrow[]{t\rightarrow\infty} \frac{1}{4}$). The concurrence
$C(\rho)=\max\bigl\{0,\frac{1}{2}(3e^{-\lambda_B t}-1)\bigr\}$ decreases exponentially and the
initially maximally entangled state ($C=1$) reaches the border of separability ($C=0$) at finite
time $t=\frac{\ln 3}{\lambda_B}$ where the mixing has the value of $\delta=\frac{1}{3}$ (see also
Ref.\cite{YuEberly}). In Fig.\ref{fig.comparison} the dependence of $\delta$ and $C$ with respect
to $\lambda t$ is shown.

\section{Realization of decoherence modes for neutron states}\label{sect.expI}

The different decoherence modes presented in Sect.\ref{sect.deco-modes} and discussed for Bell
diagonal states in Sect.\ref{sect.bell-diag} can be tested within neutron interferometry
\cite{RauchWerner}. A neutron is entangled
\cite{HasegawaLoidlBadurekBaronRauch,HasegawaLoidlBadurekBaronRauch2004,BertlmannDurstbergerHasegawaHiesmayr2004}
between the internal degree of freedom -- spin -- and the external degree of freedom -- path --
which is described by the bipartite Hilbertspace $\mathcal{H}=\mathcal{H}_{\rm
spin}\otimes\mathcal{H}_{\rm path}$. Let us consider the antisymmetric Bell state which is
experimentally feasible
\begin{equation}\label{init.state.exp}
    \lvert\Psi_{\rm exp}\rangle\equiv\lvert\Psi_4\rangle=\frac{1}{\sqrt{2}}(\lvert \Uparrow\rangle\otimes
    \lvert {\rm II}\rangle-
    \lvert \Downarrow\rangle\otimes\lvert {\rm I}\rangle)=
    \frac{1}{\sqrt{2}}(\lvert e_2\rangle-\lvert e_3\rangle)\;,
\end{equation}
where $\lvert\Uparrow\rangle$ and $\lvert\Downarrow\rangle$ represent $\pm z$ polarized
spin states whereas $\lvert{\rm I}\rangle$ and $\lvert{\rm II}\rangle$ denote the paths in the
interferometer.

Calculating the density matrices for both modes A and B we find
\begin{equation}\label{theor.A-end}
    \rho^A(t)=\frac{1}{2}\begin{pmatrix}
      0 & 0 & 0 & 0 \\
      0 & 1 & -e^{-\lambda_A t} & 0 \\
      0 & -e^{-\lambda_A t} & 1 & 0 \\
      0 & 0 & 0 & 0 \\
    \end{pmatrix}\;,
\end{equation}
\begin{equation}\label{theor.B-end}
    \rho^B(t)=\frac{1}{4}
    \begin{pmatrix}
      1-e^{-\lambda_B t} & 0 & 0 & 0 \\
      0 & 1+e^{-\lambda_B t} & -2e^{-\lambda_B t} & 0 \\
      0 & -2e^{-\lambda_B t} & 1+e^{-\lambda_B t} & 0 \\
      0 & 0 & 0 & 1-e^{-\lambda_B t} \\
    \end{pmatrix}\;.
\end{equation}
The off-diagonal elements fade away exponentially for both modes. For mode B the diagonal elements
are distributed in the whole 4-dimensional space so that at $t\rightarrow \infty$ the density
matrix approaches the normed unity, i.e., the totally mixed state.

Within neutron interferometry all matrix elements can be determined experimentally
\cite{HasegawaLoidlKleppFilippRauch} via the procedure of quantum state tomography
\cite{JamesKwiatMunroWhite}.

\subsection{Decoherence via random magnetic fields}

For the implementation of decoherence we use randomly fluctuating magnetic fields which act on an
ensemble of neutrons produced in the specific state $\rho$.

The action of a magnetic field $\vec B=B\vec n$ in the direction $\vec n$ on a neutron state is
described by the unitary operator $U(\alpha)=e^{i\frac{\alpha}{2}\vec n\cdot\vec\sigma}$, where
$\alpha=2\mu_B B t=\omega_L t$ denotes the rotation angle and $\mu_B$, $\omega_L$ the Bohr
magneton and Larmor frequency, respectively.

The neutron beam passes a fluctuating magnetic field in such a way that each neutron which is part
of the quantum mechanical ensemble described by $\rho$ feels separately a different but constant
magnetic field. This corresponds to applying a unitary operator $U(\alpha)$ with constant rotation
angle $\alpha$ onto the density matrix $\rho$. For the whole ensemble we have to take the integral
over all possible rotation angles $\alpha$
\begin{equation}
    \rho\longrightarrow\rho'=\int \underbrace{U(\alpha)\,\rho\,
    U^\dag(\alpha)}_{\rho(\alpha)}\;P(\alpha)d\alpha\;,
\end{equation}
where $P(\alpha)$ denotes a distribution function. In our case the distribution function is a
Gaussian $P(\alpha)=\frac{1}{\sqrt{2\pi}\sigma}e^{-\frac{\alpha^2}{2\sigma^2}}$ with standard
deviation $\sigma$. Although each transformation separately is unitary due to the integration we
end up with a nonunitary evolution.

\subsection{Mode A}

\begin{figure}[htbp]
\center\rule{-2cm}{0cm}
  \begin{pspicture}(0,-0.5)(8,5.3)
\psline(0.5,4)(7,4)
\psline(2,4)(2,1)
\psline(2,1)(9.8,1)
\psline(7,4)(7,0)
\psline(1.75,4.25)(2.25,3.75)
\psline(6.77,1.23)(7.25,0.75)
\psline(6.75,4.25)(7.25,3.75)
\psline(1.75,1.25)(2.25,0.75)
\rput(1,4.4){$\lvert \Downarrow\rangle$}
\rput(3,4.4){$\lvert{\rm I}\rangle$}
\rput(1.5,3){$\lvert{\rm II}\rangle$}
\pspolygon[fillcolor=black,fillstyle=solid](1.7,1.9)(2.3,1.9)(2.3,2.1)(1.7,2.1)
\rput(0.5,2){Spin flipper}
\pspolygon[fillcolor=lightgray,fillstyle=solid](4.5,3.5)(5,3.5)(5,4.5)(4.5,4.5)
\rput(4.8,4.8){$B_z^{({\rm I})}(\alpha)$}
\pspolygon[fillcolor=lightgray,fillstyle=solid](4.1,0.5)(4.6,0.5)(4.6,1.5)(4.1,1.5)
\rput(4.45,0){$B_z^{({\rm II})}(\beta)$}
\pswedge(7,-0.2){0.3}{180}{360}
\pswedge(10,1){0.3}{270}{90}
\pspolygon[fillcolor=lightgray,fillstyle=solid](6,0.6)(7.5,2.1)(7.4,2.2)(5.9,0.7)
\rput(9,2){\small{Phase shifter $\chi$}}
\pspolygon[fillcolor=lightgray,fillstyle=solid](8.3,0.75)(8.5,0.75)(8.5,1.25)(8.3,1.25)
\rput(8.5,0.5){\small{Spin rotator $\xi$}}
\end{pspicture}\\
  \caption{Experimental setup for the realization of mode A. The spin flipper is inserted
  to achieve the Bell singlet state. The magnetic fields $B_z^{({\rm I})}(\alpha)$ and $B_z^{({\rm II})}(\beta)$ produce
  independent rotations $U(\alpha)$ and $U(\beta)$, respectively. With the phase shifter $\chi$ and the
  spin rotator $\xi$ the final state is analyzed.}\label{setupA}
\end{figure}
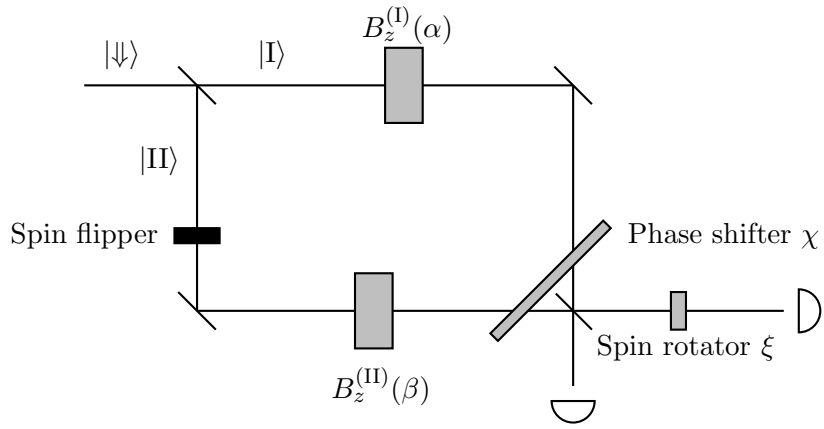

For an incoming polarized neutron the state $\lvert\Psi_{\rm exp}\rangle$ is prepared after
passing the beam splitter and spin flipper. This initial state is subjected to the fluctuating
magnetic fields oriented along the $z$-axis in each path of the interferometer, see
Fig.\ref{setupA}. The rotations $U(\alpha)$ and $U(\beta)$ caused by the fields are independent
but their distributions have the same deviation $\sigma$.

The action of the two magnetic fields can be described by a ``conditioned operation''. Depending
on the state of the spatial degree of freedom either operation $U(\alpha)$ or $U(\beta)$ is
applied to the spin state
\begin{equation}
\begin{split}
    \lvert \psi_{\rm spin}\rangle\otimes\lvert{\rm I}\rangle&\longrightarrow
    U(\alpha)\lvert \psi_{\rm spin}\rangle\otimes\lvert{\rm I}\rangle\\
    \lvert \psi_{\rm spin}\rangle\otimes\lvert{\rm II}\rangle&\longrightarrow
    U(\beta)\lvert \psi_{\rm spin}\rangle\otimes\lvert{\rm II}\rangle\;.
\end{split}
\end{equation}
For a single neutron the application of the conditioned operation on the initial state
$\lvert\Psi_{\rm exp}\rangle$, Eq.\eqref{init.state.exp}, gives
\begin{equation}
    \rho(\alpha,\beta)=\frac{1}{2}\begin{pmatrix}
      0 & 0 & 0 & 0 \\
      0 & 1 & -e^{i\frac{\alpha+\beta}{2}} & 0 \\
      0 & -e^{-i\frac{\alpha+\beta}{2}} & 1 & 0 \\
      0 & 0 & 0 & 0 \\
    \end{pmatrix}\;,
\end{equation}
which after integration over $\alpha$ and $\beta$ turns into
\begin{equation}\label{exp.A-end}
    \rho'=\int \rho(\alpha,\beta)\,P(\alpha)\,P(\beta)d\alpha\, d\beta=
    \frac{1}{2}\begin{pmatrix}
      0 & 0 & 0 & 0 \\
      0 & 1 & -e^{-\frac{\sigma^2}{4}} & 0 \\
      0 & -e^{-\frac{\sigma^2}{4}} & 1 & 0 \\
      0  & 0 & 0 & 0 \\
    \end{pmatrix}\;.
\end{equation}
By comparison of Eq.\eqref{theor.A-end} and Eq.\eqref{exp.A-end} we immediately see that
\begin{equation}\label{conI}
    \lambda_A t=\frac{\sigma^2}{4}\;,
\end{equation}
the decoherence parameter $\lambda_A$ is directly related to the deviation $\sigma$ of the
fluctuating magnetic fields. Note, that for only one magnetic field located in one of the paths
fluctuating with deviation $\sigma$ the above relation is given by $\lambda_A
t=\frac{\sigma^2}{8}$, and for one field acting on both paths we have $\lambda_A
t=\frac{\sigma^2}{2}$.

\subsection{Mode B}

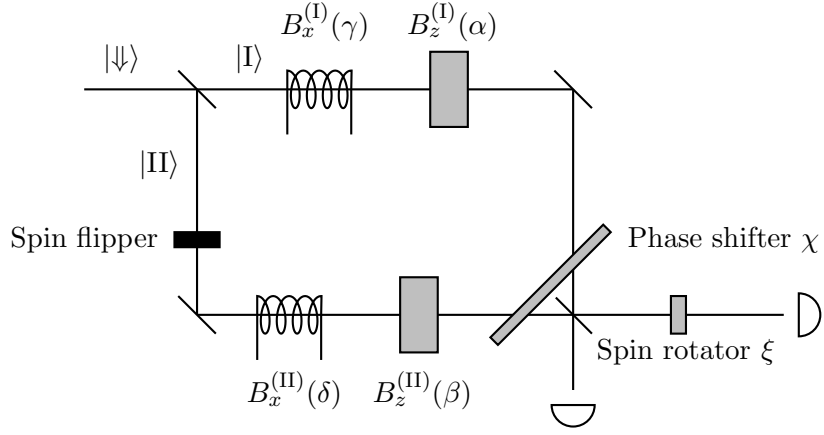
\begin{figure}[htbp]
\center\rule{-2cm}{0cm}
  \begin{pspicture}(0,-0.5)(8,5.3)
\psline(0.5,4)(7,4)
\psline(2,4)(2,1)
\psline(2,1)(9.8,1)
\psline(7,4)(7,0)
\psline(1.75,4.25)(2.25,3.75)
\psline(6.77,1.23)(7.25,0.75)
\psline(6.75,4.25)(7.25,3.75)
\psline(1.75,1.25)(2.25,0.75)
\rput(1,4.4){$\lvert \Downarrow\rangle$}
\rput(2.7,4.4){$\lvert{\rm I}\rangle$}
\rput(1.5,3){$\lvert{\rm II}\rangle$}
\pspolygon[fillcolor=black,fillstyle=solid](1.7,1.9)(2.3,1.9)(2.3,2.1)(1.7,2.1)
\rput(0.5,2){Spin flipper}
\pscoil[coilwidth=0.5,coilheight=0.6,coilarm=0](3.2,4)(4.05,4)
\psline(3.2,4)(3.2,3.4)
\psline(4.05,4)(4.05,3.4)
\rput(3.7,4.9){$B_x^{({\rm I})}(\gamma)$}
\pspolygon[fillcolor=lightgray,fillstyle=solid](5.1,3.5)(5.6,3.5)(5.6,4.5)(5.1,4.5)
\rput(5.4,4.9){$B_z^{({\rm I})}(\alpha)$}
\pscoil[coilwidth=0.5,coilheight=0.6,coilarm=0](2.8,1)(3.65,1)
\psline(2.8,1)(2.8,0.4)
\psline(3.65,1)(3.65,0.4)
\rput(3.3,0){$B_x^{({\rm II})}(\delta)$}
\pspolygon[fillcolor=lightgray,fillstyle=solid](4.7,0.5)(5.2,0.5)(5.2,1.5)(4.7,1.5)
\rput(5.0,0){$B_z^{({\rm II})}(\beta)$}
\pswedge(7,-0.2){0.3}{180}{360}
\pswedge(10,1){0.3}{270}{90}
\pspolygon[fillcolor=lightgray,fillstyle=solid](6,0.6)(7.5,2.1)(7.4,2.2)(5.9,0.7)
\rput(9,2){\small{Phase shifter $\chi$}}
\pspolygon[fillcolor=lightgray,fillstyle=solid](8.3,0.75)(8.5,0.75)(8.5,1.25)(8.3,1.25)
\rput(8.5,0.5){\small{Spin rotator $\xi$}}
\end{pspicture}\\
  \caption{Experimental setup for the realization of mode B. The magnetic fields $B_x^{({\rm I})}(\gamma)$,
  $B_x^{({\rm II})}(\delta)$, $B_z^{({\rm I})}(\alpha)$ and $B_z^{({\rm II})}(\beta)$ generate
  independent unitary rotations $U(\gamma)$, $U(\delta)$, $U(\alpha)$ and $U(\beta)$, respectively.
  The order of the magnetic fields in each path does not matter in this context.}\label{setupB}
\end{figure}

For mode B we prepare the same state $\lvert\Psi_{\rm exp}\rangle$ but use different fluctuating
magnetic fields, shown in Fig.\ref{setupB}. The different unitary operations caused by the
magnetic fields are assumed to act independently but the Gaussian distributions have the same
deviation $\sigma$.

In order to implement experimentally the rotated projectors \eqref{proj.states} we need a magnetic
field in $x$-direction $B_x$. However, to achieve the same damping in the off-diagonal elements as
in the diagonal elements we have to insert an additional magnetic field in $z$-direction $B_z$
which influences only the off-diagonal elements. It reflects somehow the effect of the Kraus
operators which act in $x$- and $z$-direction (see operator $M_3$ in Eq.\eqref{krausB}).

The neutron after the conditioned operation is described by the density matrix
\begin{equation}
\begin{split}
    &\rho(\alpha,\beta,\gamma,\delta)=\\
    &=\frac{1}{2}{\small \begin{pmatrix}
      \sin^2\frac{\gamma}{2} & -i\sin\frac{\gamma}{2}\cos\frac{\delta}{2} e^{i\frac{\alpha-\beta}{2}} & \frac{1}{2}i \sin\gamma e^{i\alpha} &  -\sin\frac{\gamma}{2}\sin\frac{\delta}{2} e^{i\frac{\alpha+\beta}{2}} \\
      i\sin\frac{\gamma}{2}\cos\frac{\delta}{2} e^{-i\frac{\alpha-\beta}{2}} & \cos^2\frac{\delta}{2} & -\cos\frac{\gamma}{2}\cos\frac{\delta}{2} e^{i\frac{\alpha+\beta}{2}} & -\frac{1}{2}i \sin\delta e^{i\beta} \\
      -\frac{1}{2}i \sin\gamma e^{-i\alpha} & -\cos\frac{\gamma}{2}\cos\frac{\delta}{2} e^{-i\frac{\alpha+\beta}{2}} & \cos^2\frac{\gamma}{2} & i\cos\frac{\gamma}{2}\sin\frac{\delta}{2} e^{-i\frac{\alpha-\beta}{2}} \\
      -\sin\frac{\gamma}{2}\sin\frac{\delta}{2} e^{-i\frac{\alpha+\beta}{2}} & \frac{1}{2}i \sin\delta e^{-i\beta} & -i\cos\frac{\gamma}{2}\sin\frac{\delta}{2} e^{i\frac{\alpha-\beta}{2}} & \sin^2\frac{\delta}{2}
       \\
    \end{pmatrix}}\;.
\end{split}
\end{equation}
For the ensemble state we get after the integrations over the angles $\alpha$, $\beta$, $\gamma$,
$\delta$ with Gaussian weights
\begin{equation}\label{exp.B-end}
    \rho'=\frac{1}{4}
    \begin{pmatrix}
      1-e^{-\frac{\sigma^2}{2}} & 0 & 0 & 0 \\
      0 & 1+e^{-\frac{\sigma^2}{2}} & -2e^{-\frac{\sigma^2}{2}} & 0 \\
      0 & -2e^{-\frac{\sigma^2}{2}} & 1+e^{-\frac{\sigma^2}{2}} & 0 \\
      0 & 0 & 0 & 1-e^{-\frac{\sigma^2}{2}} \\
    \end{pmatrix}\;,
\end{equation}
and by comparing Eqs.\eqref{theor.B-end} and \eqref{exp.B-end} we find the relation
\begin{equation}\label{conII}
    \lambda_B t=\frac{\sigma^2}{2}
\end{equation}
between the decoherence parameter $\lambda_B$ and the deviation $\sigma$ of the Gaussian distribution.\\

\emph{Experimental test.} Our decoherence modes A and B can be tested experimentally in the
following way. The incoming polarized neutron, prepared after the beam splitter and spin flipper
in a Bell singlet state, is subjected to magnetic fields with a certain Gaussian variation
$\sigma$ in both paths of the interferometer (see Fig.\ref{setupA} for mode A and Fig.\ref{setupB}
for mode B). Due to relation \eqref{conI} for mode A and \eqref{conII} for mode B the value of the
decoherence parameter $\lambda$ -- the ``dephasing'' due to the variation of the magnetic fields
-- is adjusted. The time $t$ corresponds to the duration the neutron remains in the
interferometer, more precisely, within the magnetic fields and remains constant. The individual
density matrix elements are measured experimentally via quantum state tomography and have to be
compared with the corresponding theoretical expressions \eqref{theor.A-end} and
\eqref{theor.B-end}. By varying $\sigma$, which means varying $\lambda$, one can nicely examine
the specific exponential decrease of the decoherence modes A and B.

\section{Connection to the Kraus operator decomposition}\label{sect.expII}

In the following section we present a connection to the Kraus operator decomposition. According to
the theory of decoherence
\cite{BreuerPetruccione,NielsenChuang,GiuliniJoosKieferKupschStamatescuZeh} the non-unitary
evolution of the system can be described by Kraus operators. We want to demonstrate that by a
simple experiment within neutron interferometry. We can check whether the theoretically predicted
Kraus operators correspond to the implemented decoherence modes discussed in Sect.\ref{sect.expI}.

\subsection{Kraus operator decomposition}

The completely positive time evolution generated by the Liouville -- von Neumann master equation
\eqref{time-evolution} together with the Lindblad form of the dissipator \eqref{dissipator-allg}
can also be represented by a dynamical map expressed in the Kraus operator decomposition
\cite{Kraus,NielsenChuang}
\begin{equation}\label{kraus-decomposition}
    \rho(0)\mapsto \rho(t)=\sum_k M_k\rho(0) M_k^{\dag}\;,
\end{equation}
where the Kraus operators $M_k$ fulfil $\sum_k M_k^{\dag}M_k=\mathbbm{1}$. The first approach
corresponds to a continuous time-dependence of the state whereas the second one treats decoherence
via discrete state changes. Both views are equivalent and a correspondence between Lindblad
generators $A_k$ and Kraus operators $M_k$ exists for small $\delta t$ (see e.g.
Ref.\cite{Preskill-notes})
\begin{equation}\label{connectin-kraus-lindblad}
\begin{split}
    M_0&=\mathbbm1-(i H+\frac{1}{2}\sum A_k^\dag A_k)\delta t\\
    M_k&=\sqrt{\delta t}\, A_k\;.
\end{split}
\end{equation}
Clearly, the Kraus operators are not uniquely determined by Eq.\eqref{kraus-decomposition} and
allow for a unitary transformation.

\subsection{Mode A}

For the Hilbert space we are using, $\mathcal{H}_{\rm spin}\otimes\mathcal{H}_{\rm path}$, where
entanglement occurs between spin and spatial degrees of freedom, mode A represents a kind of phase
flip channel \cite{NielsenChuang,Preskill-notes} which destroys the coherence of the system. In
this case the Kraus operators are given by
\begin{equation}\label{krausA}
\begin{split}
    M_0&=\sqrt{1-\frac{3w}{4}}\;\mathbbm{1}^{s}\otimes\mathbbm{1}^{p}\hspace{2cm}
    M_1=\sqrt{\frac{w}{4}}\;\mathbbm{1}^{s}\otimes\sigma_z^{p}\\
    M_2&=\sqrt{\frac{w}{4}}\;\sigma_z^{s}\otimes\mathbbm{1}^{p}\hspace{2.7cm}
    M_3=\sqrt{\frac{w}{4}}\;\sigma_z^{s}\otimes\sigma_z^{p}\;,
\end{split}
\end{equation}
where $w=\lambda t$ is the probability for the occurring decoherence. It leads for small
$\delta t$ to the state $\rho(t)$, Eq.\eqref{timeevol.4dim1}, which allows for a general
initial condition.

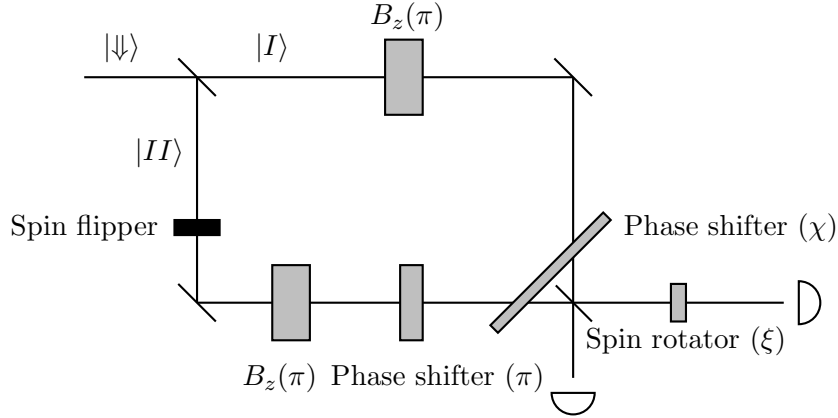
\begin{figure}[htbp]
\center\rule{-2cm}{0cm}
  \begin{pspicture}(0,-0.5)(8,5.3)
\psline(0.5,4)(7,4)
\psline(2,4)(2,1)
\psline(2,1)(9.8,1)
\psline(7,4)(7,0)
\psline(1.75,4.25)(2.25,3.75)
\psline(6.77,1.23)(7.25,0.75)
\psline(6.75,4.25)(7.25,3.75)
\psline(1.75,1.25)(2.25,0.75)
\pspolygon[fillcolor=lightgray,fillstyle=solid](4.5,3.5)(5,3.5)(5,4.5)(4.5,4.5)
\rput(4.8,4.8){$B_z(\pi)$}
\pspolygon[fillcolor=lightgray,fillstyle=solid](3,0.5)(3.5,0.5)(3.5,1.5)(3,1.5)
\rput(3.1,0){$B_z(\pi)$}
\pswedge(7,-0.2){0.3}{180}{360}
\pswedge(10,1){0.3}{270}{90}
\pspolygon[fillcolor=lightgray,fillstyle=solid](4.7,0.5)(5,0.5)(5,1.5)(4.7,1.5)
\rput(5.2,0){Phase shifter ($\pi$)}
\pspolygon[fillcolor=lightgray,fillstyle=solid](6,0.6)(7.5,2.1)(7.4,2.2)(5.9,0.7)
\rput(9.1,2){\small{Phase shifter ($\chi$)}}
\pspolygon[fillcolor=lightgray,fillstyle=solid](8.3,0.75)(8.5,0.75)(8.5,1.25)(8.3,1.25)
\rput(8.5,0.5){\small{Spin rotator ($\xi$)}}
\rput(1,4.4){$\lvert \Downarrow\rangle$}
\rput(3,4.4){$\lvert I\rangle$}
\rput(1.5,3){$\lvert II\rangle$}
\pspolygon[fillcolor=black,fillstyle=solid](1.7,1.9)(2.3,1.9)(2.3,2.1)(1.7,2.1)
\rput(0.5,2){Spin flipper}
\end{pspicture}\\
  \caption{Experimental setup for the realization of the Kraus operator $\sigma_z^s\otimes\sigma_z^p$
  for mode A.}\label{setupAII}
\end{figure}

Experimentally the Kraus operators can be implemented in the interferometer in the
following way. Incoming polarized neutrons are prepared as Bell singlet states and feel
the effect caused by the Kraus operators, as shown in Fig.\ref{setupAII} for the operator
$M_3$. The identity operators $\mathbbm{1}^{s}$ and $\mathbbm{1}^{p}$ clearly do not
change the spin and spatial degree of freedom. The operator $\sigma_z^{s}$ when acting on
the spin state $\lvert\Downarrow\rangle$ induces a phase shift of $\pi$. This phase shift
difference between spin up and spin down can be implemented by a magnetic field in
$z$-direction $B_z$ (modulo an overall phase shift). The operator $\sigma_z^{p}$ on the
spatial subspace is realized by a phase shifter in the path $\lvert{\rm II}\rangle$ which
induces a fixed phase shift of $\pi$.

The states produced by the four Kraus operators are measured tomographically and the weighted sum
according to \eqref{krausA} represents the state $\rho$, Eq.\eqref{theor.A-end}, of mode A.

\subsection{Mode B}

Mode B is a combination of a bit flip channel and a phase flip channel
\cite{NielsenChuang,Preskill-notes}. The corresponding Kraus operators are
\begin{equation}\label{krausB}
\begin{split}
    M_0&=\sqrt{1-\frac{3w}{4}}\;\mathbbm{1}^{s}\otimes\mathbbm{1}^{p}\hspace{2cm}
    M_1=\sqrt{\frac{w}{4}}\;\mathbbm{1}^{s}\otimes\sigma_z^{p}\\
    M_2&=\sqrt{\frac{w}{4}}\;\sigma_x^{s}\otimes\mathbbm{1}^{p}\hspace{2.7cm}
    M_3=\sqrt{\frac{w}{4}}\;\sigma_x^{s}\otimes\sigma_z^{p}\;,
\end{split}
\end{equation}
and create for small $\delta t$ the state given by Eqs. \eqref{timeevol.4dim2a},
\eqref{timeevol.4dim2b} and \eqref{timeevol.4dim2c} ($w=\lambda t$) allowing for general
initial conditions.

\begin{figure}[htbp]
\center\rule{-2cm}{0cm}
  \begin{pspicture}(0,-0.5)(8,5.1)
\psline(0.5,4)(7,4)
\psline(2,4)(2,1)
\psline(2,1)(9.8,1)
\psline(7,4)(7,0)
\psline(1.75,4.25)(2.25,3.75)
\psline(6.77,1.23)(7.25,0.75)
\psline(6.75,4.25)(7.25,3.75)
\psline(1.75,1.25)(2.25,0.75)
\pswedge(7,-0.2){0.3}{180}{360}
\pswedge(10,1){0.3}{270}{90}
\pscoil[coilwidth=0.5,coilheight=0.6,coilarm=0](4.3,4)(5.15,4)
\psline(4.3,4)(4.3,3.4)
\psline(5.15,4)(5.15,3.4)
\rput(4.7,3){$B_x$}
\pscoil[coilwidth=0.5,coilheight=0.6,coilarm=0](2.8,1)(3.65,1)
\psline(2.8,1)(2.8,0.4)
\psline(3.65,1)(3.65,0.4)
\rput(3.1,0){$B_x$}
\pspolygon[fillcolor=lightgray,fillstyle=solid](4.7,0.5)(5,0.5)(5,1.5)(4.7,1.5)
\rput(5.2,0){Phase shifter ($\pi$)}
\pspolygon[fillcolor=lightgray,fillstyle=solid](6,0.6)(7.5,2.1)(7.4,2.2)(5.9,0.7)
\rput(9.1,2){\small{Phase shifter ($\chi$)}}
\pspolygon[fillcolor=lightgray,fillstyle=solid](8.3,0.75)(8.5,0.75)(8.5,1.25)(8.3,1.25)
\rput(8.5,0.5){\small{Spin rotator ($\xi$)}}
\rput(1,4.4){$\lvert \Downarrow\rangle$}
\rput(3,4.4){$\lvert I\rangle$}
\rput(1.5,3){$\lvert II\rangle$}
\pspolygon[fillcolor=black,fillstyle=solid](1.7,1.9)(2.3,1.9)(2.3,2.1)(1.7,2.1)
\rput(0.5,2){Spin flipper}
\end{pspicture}\\
  \caption{Experimental setup for the realization of the Kraus operator $\sigma_x^s\otimes\sigma_z^p$
  for mode B.}\label{setupBII}
\end{figure}
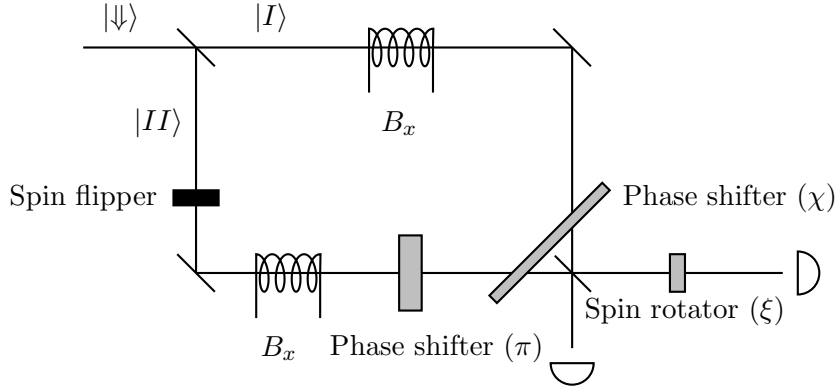

The Kraus operator for the spatial part $\sigma_z^p$ is the same as for mode A, inducing a phase
shift of $\pi$. The difference to mode A lies in the $\sigma_x^{s}$ operator for the spin part. It
can be realized by two magnetic fields $B_x$ in both arms pointing in the $x$-direction, which
cause a spin flip, see Fig.\ref{setupBII}.

Again, the weighted sum of the measured states produced by the Kraus operators according to
\eqref{krausB} leads to the state \eqref{theor.B-end} of mode B.

\section{Summary and conclusion}

We have considered the Liouville -- von Neumann equation where decoherence is implemented by the
dissipator in Lindblad form. We study two kinds of decoherence modes where the Lindblad generators
are given by different projection operators: decoherence in the eigenbasis of the Hamiltonian,
mode A, and decoherence in a rotated basis, mode B. The two modes are analyzed in detail for Bell
diagonal states, where it turns out that in mode B the state gets more mixed and the entanglement
decreases faster than in mode A. The Bell singlet state $\lvert\Psi_4\rangle$ gets separable at
finite $\lambda t=\ln 3$ in mode B whereas in mode A the state remains still entangled at that
point by an amount of $33\%$.

The realization of the proposed decoherence modes uses the bipartite Hilbert space construction of
neutron interferometry where entanglement for single neutrons occurs between an internal (spin) and
an external (path) degree of freedom.

We create decoherence via magnetic fields in the interferometer and find that the decoherence
parameter $\lambda$ is determined by the deviation $\sigma$ of the fluctuating fields, Eqs.
\eqref{conI} and \eqref{conII}. This allows an experimental control of the implemented decoherence
in each mode. The strength of decoherence does not depend on the actual rotation parameter
$\alpha$ of the magnetic field but only on the width of the Gaussian distribution.

Measuring experimentally the matrix elements of a state via state tomography and varying $\sigma$
we examine the time evolution of the state according to mode A and mode B, Eqs.
\eqref{theor.A-end} and \eqref{theor.B-end}.

In addition we can test experimentally the validity of the Kraus operator decomposition which
alternatively describes the completely positive time evolution. The Kraus operators are
constructed for each mode and realized within neutron interferometry.

\begin{acknowledgments}

The authors want to thank Stefan Filipp, J\"urgen Klepp and Helmut Rauch for helpful
discussions.

This research has been supported by the EU project EURIDICE EEC-TMR program
HPRN-CT-2002-00311 (R.A.B., K.D.), the University of Vienna (F\"orderungsstipendium of
K.D.) and the FWF-project P17803-N02 of the Austrian Science Foundation (Y.H.).

\end{acknowledgments}

\bibliographystyle{prsty}
\bibliography{bibliographie}

\end{document}